\documentclass[twocolumn]{aastex631}  
\usepackage{amsmath}
\usepackage{graphicx}
\usepackage{threeparttable}
\usepackage{txfonts}

\hypersetup{allcolors=blue}
\graphicspath{{./Figures/}}
\newcommand{\gtsima}{$\; \buildrel > \over \sim \;$}
\newcommand{\ltsima}{$\; \buildrel < \over \sim \;$}
\newcommand{\prosima}{$\; \buildrel \propto \over \sim \;$}
\newcommand{\gsim}{\lower.5ex\hbox{\gtsima}}
\newcommand{\lsim}{\lower.5ex\hbox{\ltsima}}
\newcommand{\simgt}{\lower.5ex\hbox{\gtsima}}
\newcommand{\simlt}{\lower.5ex\hbox{\ltsima}}
\newcommand{\simpr}{\lower.5ex\hbox{\prosima}}
\usepackage{xcolor}
\usepackage{soul}
\definecolor{needsfix}{RGB}{255,170,210}

\usepackage[normalem]{ulem}

\usepackage[normalem]{ulem}

\begin{document}  

\title{JWST/MIRI Spectroscopy of the Quiescent Black Hole X-ray Binary XTE J1118+480: Constraints on Jet and Circumbinary Dust Contributions to the Mid-infrared Emission}

%\title{Dust over Jet: JWST/MIRI Spectroscopy of the Quiescent Black Hole X-ray Binary XTE J1118+480}
\author[0009-0000-8524-8344]{Zihao Zuo}
\affiliation{Department of Astronomy, University of Michigan, 1085 S University, Ann Arbor, MI 48109, USA}
\author[0000-0001-5802-6041]{Elena Gallo}
\affiliation{Department of Astronomy, University of Michigan, 1085 S University, Ann Arbor, MI 48109, USA}
\author[0000-0002-3950-5386]{Nuria Calvet}
\affiliation{Department of Astronomy, University of Michigan, 1085 S University, Ann Arbor, MI 48109, USA}
\author[0000-0002-7092-0326]{Richard M. Plotkin}
\affiliation{Physics Department, University of Nevada, Reno, 1664 N. Virginia St, Reno, NV, 89557, USA}
\affiliation{Nevada Center for Astrophysics, University of Nevada, Las Vegas, NV 89154, USA}
\author[0000-0002-3500-631X]{David M. Russell}
\affiliation{Center for Astrophysics and Space Science (CASS), New York University Abu Dhabi, PO Box 129188, Abu Dhabi, UAE}
\author[0000-0001-7255-3251]{Gabriele Cugno}
\affiliation{Department of Astrophysics, University of Zurich, Winterthurerstrasse 190, 8057 Zurich, Switzerland}
\affiliation{Center for Astrophysics, Harvard \& Smithsonian, 60 Garden Street, Cambridge, MA, USA 02138}
\author[0000-0002-5654-2744]{Rob Fender}
\affiliation{Astrophysics, Department of Physics, University of Oxford, Keble Road, Oxford OX1 3RH, UK}
\affiliation{Department of Astronomy, University of Cape Town, Private Bag X3, Rondebosch 7701, South Africa}
\author[0000-0003-3352-2334]{Fraser Lewis}
\affiliation{Faulkes Telescope Project, Cardiff, Wales, UK}
\affiliation{The Schools' Observatory, Astrophysics Research Institute, Liverpool John Moores University, 146 Brownlow Hill, Liverpool L3 5RF, UK}
\author[0000-0003-3503-3446]{Joseph Michail}
\altaffiliation{NSF Astronomy \& Astrophysics Postdoctoral Fellow}
\affiliation{Center for Astrophysics, Harvard \& Smithsonian, 60 Garden Street, Cambridge, MA, USA 02138}
\author[0000-0003-2869-7682]{Jon M. Miller}
\affiliation{Department of Astronomy, University of Michigan, 1085 S University, Ann Arbor, MI 48109, USA}
\author[0000-0003-3124-2814]{James C.A. Miller-Jones}
\affiliation{International Centre for Radio Astronomy Research, Curtin University, GPO Box U1987, Perth, WA 6845, Australia}
\author[0000-0002-4436-6923]{Fan Zou}
\affiliation{Department of Astronomy, University of Michigan, 1085 S University, Ann Arbor, MI 48109, USA}

\begin{abstract}
We report on JWST/MIRI low-resolution spectroscopy ($5$--$12~\mu$m) and imaging
($15$--$25~\mu$m) of the black hole X-ray binary XTE~J1118+480 in quiescence, at an X-ray
luminosity nine orders of magnitude sub-Eddington, together with contemporaneous optical monitoring
and a radio upper limit of $4.5~\mu$Jy at 6~GHz.
With the donor subtracted, the $5$--$12~\mu$m continuum is curved,
turning over near $6~\mu$m; a single power law is rejected in favor of a $764\pm9$~K blackbody.
Longward of the turnover the spectral index is $\alpha = +1.21\pm0.36$, steeper at the $2\sigma$
level than the flat to slightly inverted spectrum of a compact jet.
Interpreted as synchrotron, the turnover implies equipartition fields of $\sim10^{4}$~G on
scales of tens of gravitational radii, but the MIRI photometry at $15$--$21~\mu$m--taken $2.6$~hr
before the spectrum--lies $30$--$90$ per cent above its extrapolation, suggesting a second
self-absorbed zone. A single partially self-absorbed jet spectrum can reach the photometry only at
the expense of the spectral curvature and the $25.5~\mu$m upper limit, so a steady compact jet is
disfavored, although not excluded. Representative self-consistent irradiated-disk models instead
reproduce the principal features of the $5$--$25~\mu$m spectral energy distribution, with a
directly illuminated dust wall at $\simeq5$ binary separations, outside the tidal-truncation
radius, and a minimum dust mass of order $10^{22}$~g.
This is the first spectroscopic evidence, though not yet conclusive, favoring circumbinary dust
over a jet in a quiescent black hole X-ray binary.
\end{abstract}

%%%%%%%%%%%%%%%%%%%%%

\section{Introduction}
\label{sec:intro}

Black hole low-mass X-ray binaries (BH-LMXBs) accrete from a Roche-lobe-filling donor of
$\lesssim1\,M_\odot$ and spend most of their lives in a faint, quiescent state
($L\lesssim10^{-5}L_\mathrm{Edd}$). In the hard state of their outbursts their radio to mid-infrared emission is dominated by a
compact, partially self-absorbed jet with a flat-to-inverted spectrum ($F_\nu\propto\nu^\alpha$,
$\alpha\approx0$--$0.5$; \citealt{Fender2001}), in which synchrotron radiation from many self-absorbed zones
at progressively larger distances from the base is superposed. This compact jet is thought to persist
into quiescence owing to radio detections in a handful of systems
\citep{gallo2006, Plotkin2021, dePolo2022}, although both the accretion flow and jet remain poorly
constrained at the lowest accretion rates \citep{Remillard2006, Done2007, Plotkin2015}.

The wavelength at which the quiescent radio jet becomes optically thin is the critical unknown.
If the partially self-absorbed spectrum extends into the infrared, the integrated jet power exceeds that dissipated
locally within the X-ray emitting accretion flow \citep{Fender2003, gallo2006, Gallo2007}, which
would overturn the classical picture in which accreting black holes radiate most of the
dissipated power within the inflow. Repeated multiwavelength studies place the break in the
near-infrared for luminous hard-state systems \citep{Russell2013}; in quiescence it has been constrained only in Swift~J1357.2$-$0933 \citep{Plotkin2016, Russell2018}. 

Until the Mid-Infrared Instrument (MIRI) on board the \textit{James Webb Space Telescope} (JWST),
mid-infrared spectroscopy remained largely unavailable for X-ray binaries, although many had
been detected photometrically with \textit{Spitzer}, \textit{WISE} and VLT/VISIR (see the
compilation of \citealt{GandhiGX339}). In the hard state, photometry has established ubiquitous
mid-infrared jet emission \citep{John2024}, a jet break that moves on hour time-scales
\citep{Gandhi2011}, fading and recovery across state transitions \citep{Russell2013b}, and
high-amplitude variability on minute time-scales \citep{Baglio2018}. The first MIRI results show no single
dominant mechanism in X-ray binaries: free--free emission from an obscuring wind in GRS~1915$+$105
\citep{Gandhi2025}, following the earlier detection of compact jet synchrotron in Cygnus~X-1
\citep{Rahoui2011}; synchrotron from the electrons that produce the high-energy Comptonized tail, with any compact jet
suppressed by a factor $\gtrsim300$, in the soft state of GX~339$-$4 \citep{GandhiGX339}; jet synchrotron in V404~Cygni
\citep{Borowski2025}. However, these are all comparatively luminous systems; even V404~Cygni, the most luminous
quiescent black hole binary, sits at $L_{\rm X}\sim10^{-7}$--$10^{-6}\,L_{\rm Edd}$
\citep{Bernardini2014}, two to three orders of magnitude above XTE~J1118+480. 

Only in deep quiescence can thermal emission from cooler, more distant material compete with the jet emission (if the jet indeed persists) at mid-infrared wavelengths. This is to be expected; circumbinary disks have been invoked as a sink of
orbital angular momentum shaping the evolution of compact binaries, first for cataclysmic
variables \citep{Taam2001, Taam2003} and later to account for the anomalously rapid
orbital-period decay measured in XTE~J1118+480 and A0620$-$00 \citep{GonzalezHernandez2014,
ChenLi2015}. The first evidence that such material might be present came from \textit{Spitzer}:
\citet{Muno2006} detected mid-infrared excesses above the donor photosphere in both systems and
fit them with cool blackbodies whose implied emitting radii exceed the binary separation
(see also \citealt{Wang2014}).
\citet{Gallo2007} re-analysed the same data, derived larger uncertainties on the $24~\mu$m
flux densities, and showed that partially self-absorbed jet synchrotron equally reproduces the data. 
%With two photometric points and a shallow upper limit the two could not be distinguished.

JWST program GO~3832 was designed to settle this, by observing both systems. The A0620$-$00 results \citep{Zuo2025} were somewhat unexpected: its stellar-subtracted spectrum follows a power law, varies by
tens of per cent on minute time-scales, and shows a blend of hydrogen recombination lines near
$7.5~\mu$m, which \citet{Zuo2025} attribute to thermal bremsstrahlung from an outflowing wind.

XTE~J1118+480 is the companion target and the subject of this Letter. It comprises a black hole of $\approx7\,M_\odot$
accreting from a K7V--M0V donor in a $4.078$~hr orbit at
$1.72\pm0.10$~kpc \citep{Cherepashchuk2019}. Its high Galactic latitude leaves the sight line almost
unreddened ($A_V\simeq0.05$), and its short orbital period places
the expected tidally truncated inner rim of any circumbinary dust at only a few solar radii. Section~\ref{sec:obs} describes the
observations and the donor subtraction; Section~\ref{sec:continuum} characterizes the
mid-infrared continuum and models it, first with self-absorbed synchrotron
(\S\ref{sec:synch}) and then with an irradiated circumbinary disk (\S\ref{sec:diad}).
%***********
\section{Observations, Data Reduction, and Analysis}
\label{sec:obs}

\subsection{JWST MIRI}
XTE~J1118+480 (hereafter XTE~J1118;
R.A.\ $= 11^{\mathrm h}18^{\mathrm m}10.7651^{\mathrm s}$,
Dec.\ $= +48^\circ02'12.21''$) was observed on 2024 April 11--12 UT with MIRI in both
Imaging (F1500W, F1800W, F2100W, F2550W) and Low-Resolution Spectroscopy (LRS) modes, under
program GO~3832 (PI: Gallo). The observations are summarized in Table~\ref{tab:miri_obs}, and the
reduction is described in Appendix~\ref{app:reduction}. The source is detected with
$\mathrm{S/N}$ of $11$, $13$, and $5$ in F1500W, F1800W, and F2100W, respectively, and is
undetected in F2550W, for which we report a $3\sigma$ upper limit of $13~\mu$Jy.

\subsubsection{Mid-infrared variability}
\label{sec:miri_var}
Processing the individual dithers separately (Appendix~\ref{app:reduction}), we find
marginal evidence for variation between dithers in F1500W ($2.0\sigma$) and F1800W
($3.0\sigma$), with fractional dither-to-dither rms of $\simeq22\%$, $26\%$, and $12\%$ at
F1500W, F1800W, and F2100W; the F2100W groups are consistent with a constant flux density. The same analysis applied to all field sources returns median
normalized flux densities consistent across dithers, so the variation is not driven by global
photometric systematics. The evidence for genuine photometric variability is therefore
marginal. 

The LRS integrations show approximately common-mode fluctuations of $9.9\%$, $10.4\%$, and
$12.8\%$ (rms) in the $5$--$7$, $7$--$9$, and $9$--$11~\mu$m bands. These data were obtained in
fixed-slit mode, which is not the configuration dedicated to high-precision time-series
spectrophotometry \citep{Bouwman2023}. Applying the same per-integration analysis to
$\sim100$ photometrically constant reference stars observed in the same mode shows that the
scatter is a smooth function of flux density alone, tracing a background-limited noise floor:
white dwarfs of the same flux density ($0.04$--$0.05$~mJy) show $11$--$14\%$, while
A0620$-$00 reduced identically lies well above the relation. The LRS
continuum of XTE~J1118+480 is therefore steady to the noise floor.

Because the imaging and the LRS are not simultaneous, we conservatively add the
dither-to-dither scatter in quadrature to the random-aperture error,
$\sigma_F = (\sigma_{\rm aper}^2 + \sigma_{\rm dither}^2)^{1/2}$, whenever the photometry is
compared to the spectrum. 
This term dominates $\sigma_F$ at F1500W and F1800W, and is
comparable to the random-aperture error at F2100W.

%***********
\subsection{Karl G. Jansky Very Large Array}
\label{sec:radio_red}
 XTE~J1118+480 was observed on 2024 April 12 from 00:30 -- 08:00 UT with the Karl G. Jansky Very Large Array (VLA).  The source was not detected in the radio, with a $3\sigma$ upper limit of $4.5\,\mu$Jy bm$^{-1}$ at 6~GHz.  The data reduction is described in Appendix \ref{app:vlareduction}.

%***********
\subsection{Donor-star Modeling and Stellar Subtraction}
\label{sec:stellar_sub}

The donor was first identified as a K7V/M0V star by \citet{Wagner2001}. We adopt the parameters of
\citet{Cherepashchuk2019} ($T = 4120$~K, $M = 0.1\text{--}0.2~M_\odot$, $R = 0.144\,a_0$,
$\log g \approx 4.5$) and a BT-Settl template at $T = 4100$~K, $\log g = 4.5$, solar
metallicity.

Ground-based $i'$ monitoring, with the Las Cumbres Observatory (LCO) network (PI: Russell; see Appendix \ref{app:lco}), covering the JWST epochs is fitted with a Fourier
model of the ellipsoidal modulation, phased on the \citet{Cherepashchuk2019} ephemeris
(Fig.~\ref{fig:donor_lc}); the best-fitting curve is evaluated at the orbital phase of
each individual exposure and used to scale the stellar template to $91\%$ of the observed $i'$ flux density, allowing for the non-stellar contribution discussed below, which is then averaged over each filter
bandwidth and LRS resolution element and subtracted (Appendix~\ref{app:reduction}). Because the
imaging and the LRS fall on opposite halves of the ellipsoidal cycle (phase $0.035$--$0.252$ and
$0.582$--$0.756$ respectively), each is normalized at its own phase rather than with a common
factor. The same-day monitoring spans two days and twelve orbital cycles, so the curve is well
determined over the full phase range; nine of those points fall within the LRS exposure itself
(mean $i' = 18.938\pm0.022$) and confirm the normalization at the spectroscopic phase. The long-term LCO monitoring shows no sustained brightening of the kind that defines the
active states of A0620$-$00 \citep{Cantrell2008} (Appendix~\ref{app:lco}), so we adopt the mean
ellipsoidal curve as the anchor rather than a lower envelope; any further non-stellar contribution
to the optical makes our donor an upper limit and the mid-infrared excess a lower limit.

As a check on the normalization, we compared the template with long-term LCO photometry in
$B$ through $z'$, de-reddened for $A_V=0.05$ (Figure~\ref{fig:optical_sed}). The observed flux density
rises above the template toward the blue, as expected from the $\sim13\,000$~K optical/UV
component identified by \citet{McClintock2003}, plausibly the gas-stream impact region, together
with a viscously heated disk with $\dot M\simeq4\times10^{13}$~g~s$^{-1}$ \citep{Muno2006}. With
its normalization left free, the optical SED prefers a donor share of $84\%$ of the $i'$ flux density,
so the donor we subtract is if anything slightly too bright and the excess conservative. The hot component is compact and, with a disk truncated at $\sim2\times10^{4}\,r_{\rm g}$,
contributes $\lesssim6\%$ of the excess at $5$--$11~\mu$m. A viscous disk extending to small
radii could instead supply the blue excess, but it fits the optical colors less well and would
radiate $\sim10^{3}$ times the observed X-ray luminosity \citep{McClintock2003}.

The dominant uncertainty is the template normalization rather than the photometric errors
on the monitoring data. We propagate the formal covariance of the Fourier coefficients, the
$10\%$ systematic flux-density error associated with the K3V--M0V spectral-type range
\citep[cf.][]{Zuo2025}, and the dispersion of the monitoring data about the fitted curve within
each window. 

%***********
\subsection{The Broadband Spectral Energy Distribution}
\label{sec:broadband_sed}

\begin{figure}
\includegraphics[width=0.95\linewidth]{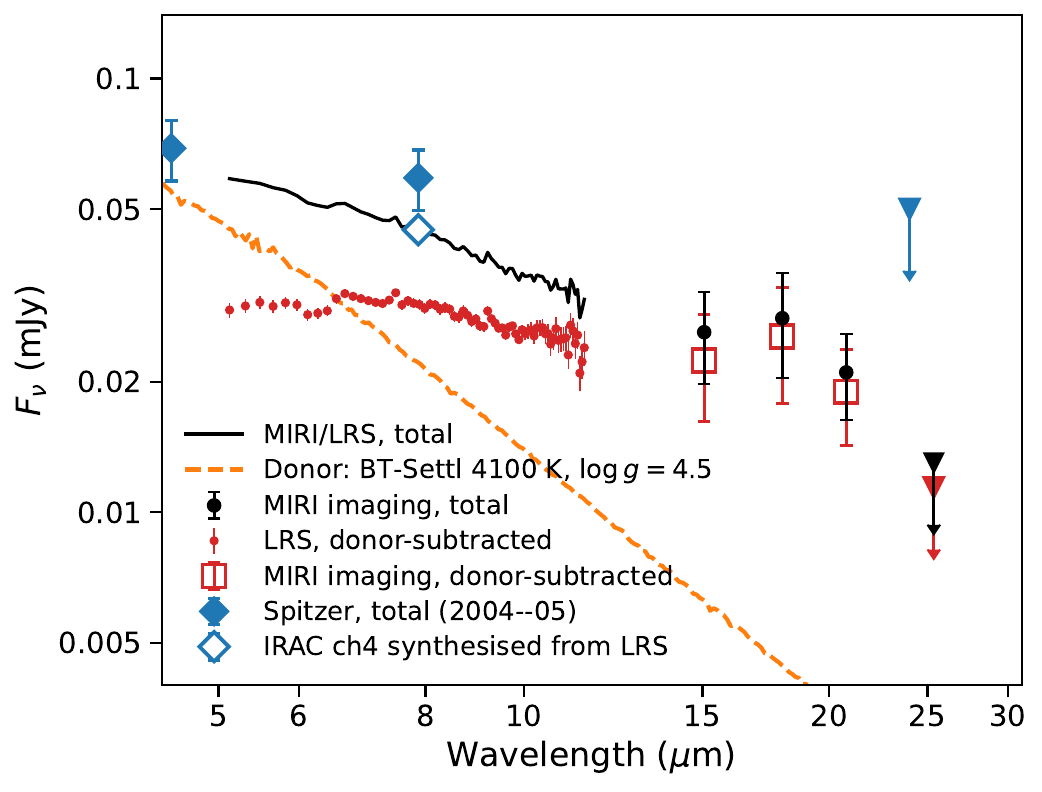}
\caption{Donor subtraction for XTE~J1118+480. Black line: the
nod-averaged MIRI/LRS spectrum before subtraction. Orange dashed: the BT-Settl donor template
($T=4100$~K, $\log g=4.5$), scaled as described in \S\ref{sec:stellar_sub} and binned to the LRS
resolution. Filled black circles: the MIRI imaging before subtraction, with the
F2550W $3\sigma$ limit (black triangle). Red circles and open squares: the donor-subtracted LRS spectrum and
imaging photometry, with the corresponding F2550W limit (red triangle). Blue diamonds: \textit{Spitzer}/IRAC
total flux densities at $4.5$ and $8~\mu$m from \citet{Gallo2007}, with the MIPS $24~\mu$m $3\sigma$
limit (blue triangle); the IRAC data were taken in 2004 November and MIPS in 2005 May, and are
plotted at the IRAC channel effective wavelengths. Error bars on the \textit{Spitzer} points are
$16\%$, the frame-to-frame scatter reported by \citet{Muno2006} for this source, rather than the
$10\%$ calibration floor adopted by \citet{Gallo2007}, since XTE~J1118+480 is the faintest object
in that sample. The open blue diamond is the IRAC channel-4 flux density synthesised from the MIRI/LRS
spectrum.}
\label{fig:donor_sub}
\end{figure}

Figure~\ref{fig:donor_sub} shows the result of the stellar subtraction. The scaled BT-Settl template accounts
fully for the optical emission but falls well short of the measured mid-infrared flux density. The ratio of donor to excess flux density is $1.06$ at $5~\mu$m,
falling to $0.47$ at $8~\mu$m and $0.28$ at $11~\mu$m, so the excess dominates over most of the
band.

For comparison, we plot the 2004 \textit{Spitzer}/IRAC and MIPS flux densities with the revised
uncertainties of \citet{Gallo2007}. At $8~\mu$m, the only
wavelength where the two overlap, \citet{Gallo2007} report $59~\mu$Jy from aperture photometry
($58~\mu$Jy from PSF fitting). Synthesising the same IRAC channel-4 measurement from our
spectrum, by integrating over $6.28$--$9.59~\mu$m gives $44.8\pm1.4~\mu$Jy. The 2024 flux density is thus some $25\%$ lower than in 2004.
Adopting the $16\%$ frame-to-frame scatter that \citet{Muno2006} quote for this source, the
difference is significant at only $1.5\sigma$ ($2.3\sigma$ with the $10\%$ calibration floor
tabulated by \citet{Gallo2007}), so a decline is suggested but not established. 
%At $24~\mu$m the \textit{Spitzer} limit of $<50~\mu$Jy is uninformative beside the $<13~\mu$Jy reached here.

%Extinction is negligible at these wavelengths for $A_V\simeq0.06$, and the donor contributes
%only a third of the $8~\mu$m flux, so the ellipsoidal modulation can move it by at most
%$\pm1.7~\mu$Jy and cannot account for the offset. A real decline would indicate variability of
%the mid-infrared emitting region on time-scales of years to decades, as expected of a dust
%structure responding to a variable irradiating luminosity.

\section{Modeling the mid-infrared excess}
\label{sec:continuum}

To characterize the excess mid-infrared emission, we first fit the stellar-subtracted LRS
spectrum ($5$--$12~\mu$m) alone, with two phenomenological forms: a single-temperature blackbody
$F_\nu^{\rm BB}\propto \nu^{3}/[\exp(h\nu/k_{\rm B}T)-1]$ and a power law $F_\nu^{\rm PL}\propto \nu^{\alpha}$.
The Gaussian-likelihood setup is identical to that used in
the analogous fit to A0620$-$00 by \citet{Zuo2025}, with broad uniform priors on the model
amplitudes (in log space), $\alpha\in[-3,3]$, and $\log_{10}(T/\mathrm{K})\in[1,4]$; parameter
uncertainties are obtained by nested sampling with \texttt{dynesty} \citep{Speagle2020}, and
models are compared by $\Delta\chi^{2}$ at equal degrees of freedom.

The LRS uncertainties returned by the pipeline underestimate the scatter between the two
independent nods by a factor of $2$--$4$, worst at the blue end, and we rescale them accordingly
(Appendix~\ref{app:reduction}). With realistic errors the blackbody is preferred over the power
law by $\Delta\chi^{2} = 71$ at equal degrees of freedom, with $\chi^{2}/{\rm dof} = 0.82$ and
$1.74$ respectively. The continuum is therefore curved and a single power law is rejected. This is the
first substantive difference from A0620$-$00, whose stellar-subtracted spectrum is well described
by a single power law over the same wavelength range \citep{Zuo2025}. The posterior blackbody
temperature is $T = 764\pm9$~K.

Curvature alone does not identify the emission mechanism: a smoothly-broken power law in the
functional form of \texttt{astropy}'s \texttt{SmoothlyBrokenPowerLaw1D}, with fixed smoothness
$\Delta=0.3$, describes the LRS equally well, with a break at
$\lambda_b\simeq6.2~\mu$m. Because this form is phenomenological, we label its indices by
wavelength rather than optical depth. The short-wavelength index is poorly constrained,
$\alpha_{\rm short}\simeq-1.5$ (68\% range $-2.4$ to $-0.8$), since only $5$--$6~\mu$m of the
spectrum lies blueward of the break.

The shape of the curvature, however, is informative. That same fit returns a long-wavelength index
$\alpha_{\rm long}$ $= +1.21\pm0.36$, steeper at the $2\sigma$ level than the $\alpha\simeq0$--$0.5$ expected of a compact jet
(and at $1.4\sigma$ than the steepest reported, $\alpha\simeq0.7$; \citealt{Russell2014,
Espinasse2018}).
%\footnote{In the Blandford--K\"onigl picture the flat-to-inverted radio spectrum arises because emission from many self-absorbed zones, each turning over at a different frequency according to its distance from the base, is superposed along the outflow, and that superposition flattens the $+5/2$ rise of any individual zone to $\alpha\simeq0$--$0.5$}
The measured index is intermediate between that and the canonical $+5/2$ of a \emph{single} homogeneous self-absorbed region.
Because it is measured red-ward of the turnover, where the donor contributes only about a third of the excess, it is also insensitive to the donor normalization (rescaling the template by $\pm20\%$, three times its uncertainty, changes
$\alpha_{\rm long}$ by less than $0.03$).

The data therefore disfavor a partially self-absorbed jet spectrum extending into the mid-infrared in XTE~J1118+480.
The spectrum is otherwise featureless. We searched for unresolved lines by fitting a
Gaussian of fixed instrumental width at every wavelength, refitting the continuum outside
each trial band. The strongest
residual is a $2.8\sigma$ excess at $7.47~\mu$m, coincident with Pf$\alpha$ and recovered
independently in both nods ($1.7\sigma$ and $2.3\sigma$). Its equivalent width is
$0.0072\pm0.0026~\mu$m, less than half the
equivalent width of the hydrogen recombination blend reported in A0620$-$00 by
\citet{Zuo2025}. We regard this as marginal, as the nod-to-nod offset in this region is
comparable to the line amplitude.

The photometry provides additional constraints. Extrapolating the LRS fits to the three photometric bands, we find that the measured flux densities lie above every extrapolation, by $30$--$90\%$
across the three bands. Folding in the full per-filter dither-to-dither rms of $12$--$26\%$ as
a conservative allowance for any change between the non-simultaneous epochs, the
combined excess persists at $2.6\sigma$ above the blackbody baseline and at $3.8\sigma$ above
the single self-absorbed synchrotron baseline. Neither is therefore an adequate description of the
$5$--$25~\mu$m spectral energy distribution.

A partially self-absorbed jet spectrum breaking near $6~\mu$m can also reach the photometry,
if the LRS and the imaging are fitted jointly. Weighting each photometric point by the number of
LRS points spanning the same interval in $\log\lambda$, a smoothly broken power law passes within
$0.7\sigma$ of all three detections, with $\alpha\simeq+0.5$ at long wavelengths, at the steep
end of the range observed for compact jets. The cost is a poorer description of the LRS
curvature ($\Delta\chi^2=29$ at the smoothness adopted above, $\simeq10$ if the smoothness is left
free) and a prediction $\simeq45\%$ above the F2550W $3\sigma$ limit ($16.5$--$17$ against
$11.4~\mu$Jy after donor subtraction). Forms that retain the LRS curvature instead fall
$0.6$--$1.7\sigma$ below each detection. The distinction therefore rests chiefly on the
$25.5~\mu$m limit, and a jet whose emission varied between the non-simultaneous imaging exposures
cannot be excluded. The marginal feature at $7.47~\mu$m, if confirmed, would instead require
thermal gas, since synchrotron produces no lines.

\begin{deluxetable*}{lccccccccccc}
\tablecaption{Summary of JWST/MIRI Imaging Observations\label{tab:miri_obs}}
\tablehead{
\colhead{Mode / Filter} &
\colhead{$\lambda_{\rm pivot}$} &
\colhead{$W_{\rm eff}$} &
\colhead{Readout / Subarray} &
\colhead{$N_{\rm grp}$} &
\colhead{$N_{\rm int}$} &
\colhead{$N_{\rm dither}$} &
\colhead{$t_{\rm tot}$ (s)} &
\colhead{Start (MJD)} &
\colhead{S/N} &
\colhead{Flux density (mJy)}
}
\startdata
F1500W & 15.06 & 2.92 & FASTR1/FULL & 8  & 3 & 3 & 66.6 & 60411.9857 & 11 &  $0.0260\pm0.0024$ \\
F1800W & 17.98 & 2.95 & FASTR1/FULL & 32 & 3  & 3 & 266.4 & 60411.9906 & 13 & $0.0280\pm0.0021$ \\
F2100W & 20.79 & 4.58 & FASTR1/FULL & 28 & 12  & 12 & 932.4 & 60411.9979 & 5 & $0.0210\pm0.0040$ \\
F2550W & 25.36 & 3.67 & FASTR1/FULL & 15 & 48  & 12 & 2097.9 & 60412.0226 & \nodata & $<0.013$ \\
LRS    & \nodata & \nodata & FASTR1/FULL & 45 & 40 & 2 & 5100.5 & 60412.0787 & \nodata & \nodata \\
\enddata
\tablecomments{
$\lambda_{\rm pivot}$ and $W_{\rm eff}$ are the pivot wavelength and effective bandwidth of each filter.
$t_{\rm tot}$ is the total on-source integration time, summed over all dithers (for the LRS, over both nods).
Quoted flux-density errors are the random-aperture uncertainties only; the larger, dither-inflated
uncertainties adopted whenever the photometry is compared with the spectrum are given in
\S\ref{sec:miri_var} and are the ones plotted in Figures~\ref{fig:donor_sub}--\ref{fig:diad}. All imaging was obtained on 2024 April 11--12 UT. The LRS exposure began $1.35$~hr after the last imaging exposure ended;
mid-exposure to the first imaging exposure is $2.6$~hr. The imaging spans orbital phase
$0.035$--$0.252$ and the LRS $0.582$--$0.756$.
}
\end{deluxetable*}

%***********
\subsection{Multi-zone synchrotron model}
\label{sec:synch}

\begin{figure}
\centering
\includegraphics[width=\linewidth]{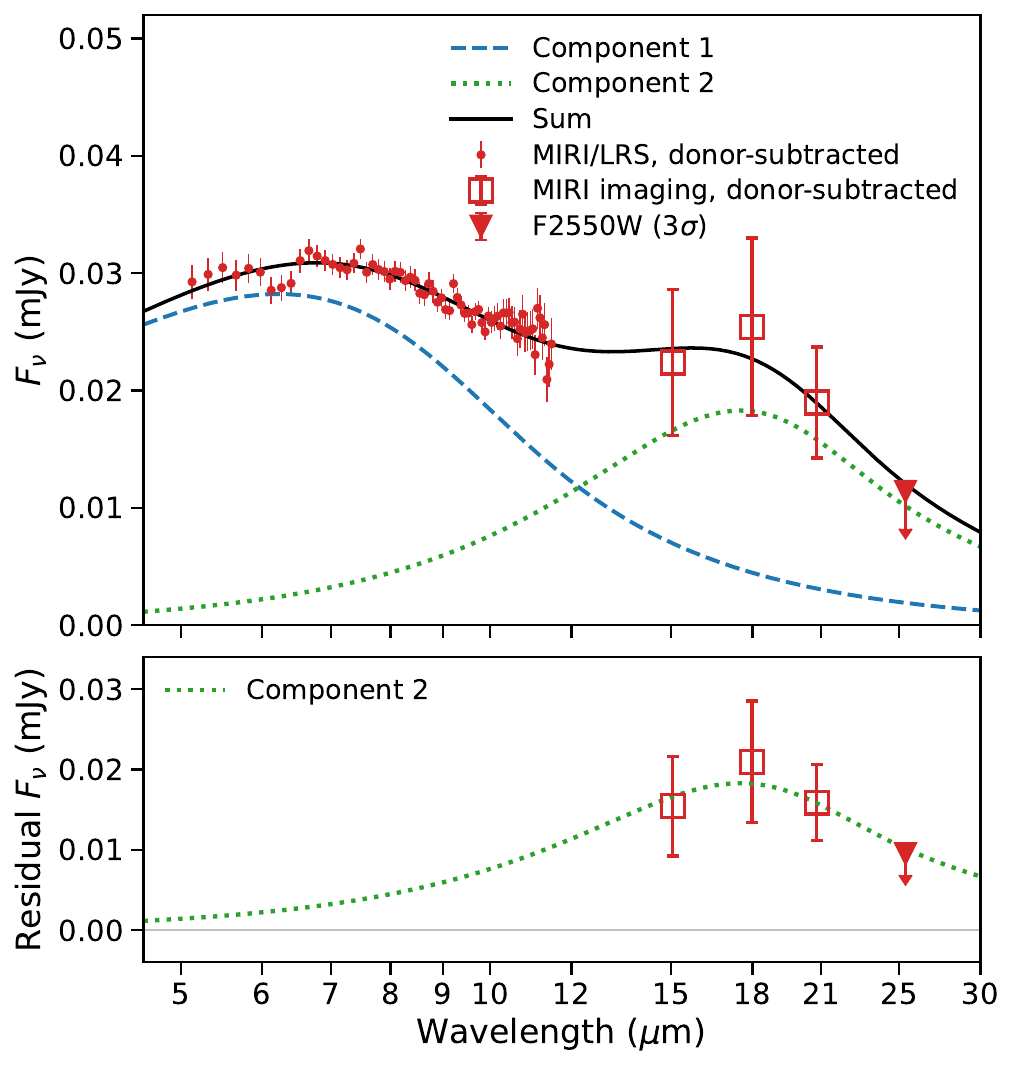}
\caption{Sequential two-zone self-absorbed synchrotron fit. \emph{Top}: the
donor-subtracted MIRI/LRS spectrum (small filled red circles, binned by three) and imaging photometry (open
red squares), with the F2550W $3\sigma$ upper limit (filled red triangle). Component~1 (blue
dashed) is fitted to the LRS, component~2 (green dotted) to the photometry after component~1 is
removed, and the two are iterated to convergence; their sum is the solid black curve.
\emph{Bottom}: the photometry after subtraction of component~1, with component~2 overlaid. }
\label{fig:seqfit}
\end{figure}

We start by fitting the stellar-subtracted LRS spectrum with a single homogeneous, self-absorbed sphere of flux density $F_\nu^{\rm SSA} = A\left(\frac{\nu}{\nu_t}\right)^{5/2}\left(1-e^{-\tau_\nu}\right)$, where $\tau_\nu = \left(\frac{\nu}{\nu_t}\right)^{-(p+4)/2}$ 
parameterized by a turnover frequency $\nu_t$ and an electron power-law index $p$ \citep{vanderLaan1966, Pacholczyk1970}. With broad uniform priors
on $p\in[1,6]$ and $\log_{10}(\nu_t/\mathrm{Hz})\in[12.5,14.5]$, the fit yields  
$\lambda_t \simeq 6.43\pm 0.30~\mu$m ($\nu_t = 4.66\times10^{13}$~Hz) and $p\simeq 1.78\pm 0.17$, with an
inferred peak flux density $S_{\rm peak}\simeq 30.6\pm 0.2~\mu$Jy.
%This parametrization describes the spectrum as well as the blackbody. 
Assuming
equipartition between the energy density in the particles and magnetic field, the standard minimum-energy relations \citep{Pacholczyk1970} return a constraint on the magnetic field strength and the sphere radius: 
$B \simeq 2.6 \times10^{4}~{\rm G}$ and 
$R_{\rm SSA} \simeq 2.0 \times10^{7}~{\rm cm}$. These correspond to $\simeq 18\,r_{\rm g}$ for $M_{\rm BH}=7\,M_\odot$, and to a
brightness temperature at the turnover of $T_{\rm b}\simeq 2\times 10^{10}$~K. A single self-absorbed synchrotron zone is therefore a physically admissible description of the
LRS continuum. However,
%A component that turns over at $9.5~\mu$m with an optically-thick $F_\nu\propto\nu^{5/2}$ branch falls below a nanojansky at 6~GHz, far beneath the $4.5~\mu$Jy limit; conversely, a flat-spectrum jet anchored at that limit could supply a substantial fraction of the mid-infrared flux. The radio is uninformative either way. 
one such component cannot account for the full spectral energy distribution, as its extrapolation redward of $12~\mu$m falls below the $15$--$21~\mu$m photometry (\S\ref{sec:continuum}). 
We therefore test whether two
self-absorbed zones can reproduce the data by fitting them sequentially rather than jointly (the $240$ LRS points
would otherwise overwhelm the four photometric ones). After fitting component~1 to the LRS alone, as described above, we evaluate
its predicted contribution at the four photometric wavelengths, subtract it, and fit component~2 to the residual.
The procedure is iterated to convergence, since component~2 contributes $6\%$ of the LRS flux density at
$5~\mu$m rising to $40\%$ at $11~\mu$m and component~1 must be refitted with that contribution
removed. The converged model reproduces the LRS to a median $1.8\%$ over
$5$--$11~\mu$m with $\chi^2/{\rm dof} = 0.81$, and is consistent, within the errors, with all three photometric detections 
(Figure~\ref{fig:seqfit}).

Following \citet{FenderBright2019}, we equate the field required by the self-absorption
condition at the turnover to the equipartition field, evaluating $\nu_{\tau=1}$ from the fitted
profile and integrating the synchrotron luminosity over $10^{12}$--$10^{15}$~Hz; changing that range
by two decades alters $B$ and $R$ by less than $30\%$, and the implementation recovers the
V404~Cyg flare parameters tabulated by \citet{FenderBright2019}. These estimates assume a uniform,
isotropic sphere without beaming, and are uncertain by factors of a few. For component 1, this gives 
$\lambda_t = 6.2~\mu$m, $p = 2.8$, 
$R_{\rm SSA} \simeq 1.5\times 10^{7}$ cm ($\simeq 14\,r_{\rm g}$), and $B \simeq 3.5\times10^{4}$ G. 
Component 2 is somewhat less constrained, with $\lambda_t = 17.5$ $\mu$m, 
$R_{\rm SSA} \simeq 3.2-4.1\times 10^{7}$ cm ($ \simeq 29-37\,r_{\rm g}$), $B \simeq 1.1-1.3\times10^{4}$ G, and $p$ spanning the entire range between 2$-$3.5, which the fit does not constrain. 
The
brightness temperature at the turnover is $T_{\rm b}\simeq2\times10^{10}$~K for both components. This configuration reproduces the SED, and the inferred parameters are ordered in the
expected sense: the component turning over at longer wavelength is the larger and the more weakly
magnetized, as self-absorption scaling requires. Both lie on the trend of magnetic field strength against emitting-region size defined by mid-infrared
jet-base estimates \citep{Gandhi2011} at one extreme and resolved radio ejecta at the other, though we caution that those are inferred for much more luminous systems. 

Because the LRS and the photometry were obtained at different times, the two components might
be one sphere observed twice. The imaging, however, preceded the LRS, so a single sphere would
have to evolve from a turnover at $17.5~\mu$m to one at $6.2~\mu$m, blueward, whereas adiabatic
expansion moves the turnover redward with time. That interpretation is therefore excluded.
Regardless, two distinct synchrotron emitting zones with no radio
counterpart and no detectable variability in either epoch (\S\ref{sec:miri_var}) describe neither
a continuous compact jet, whose superposed zones would give $\alpha\simeq0$--$0.5$, nor discrete
ejecta, which would be transient.
We therefore regard a two-component synchrotron origin as permitted albeit unlikely, and turn to the
thermal alternative.

\begin{figure*}
\centering
\includegraphics[width=\textwidth]{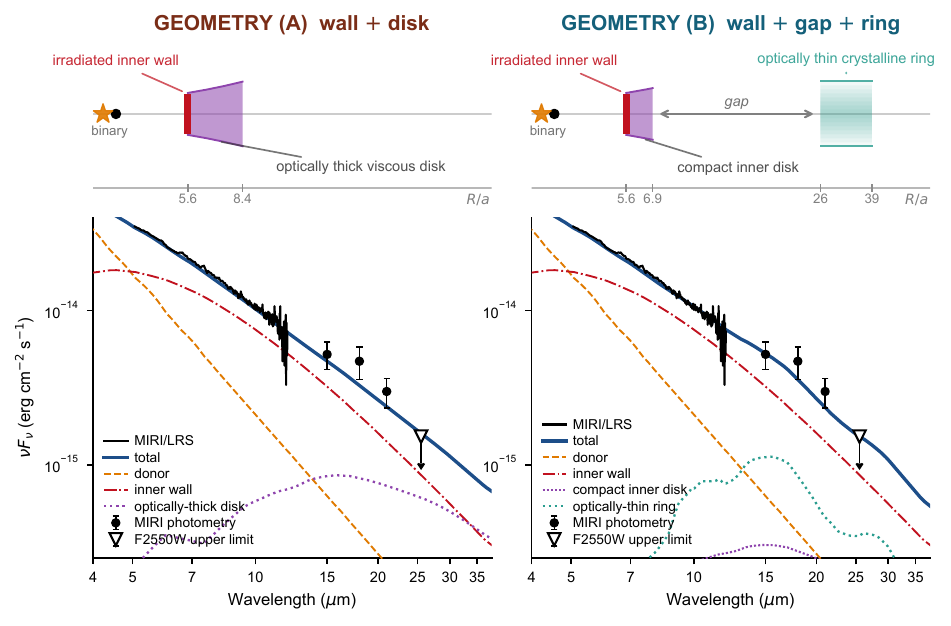}
\caption{Representative DIAD circumbinary-disk models for XTE~J1118+480
(Table~\ref{tab:diad}), plotted as $\nu F_\nu$ versus wavelength.
The upper panels sketch the assumed geometry on a logarithmic radial axis in units of the
binary separation $a$, with each component drawn in the same colour as its contribution to the
spectrum below; the vertical scale is exaggerated.
\emph{Left\ column}, geometry~(A), wall
$+$ disk: a compact, optically-thick viscous disk (magenta dotted) fills the region immediately
outside the inner wall, with no gap. \emph{Right\ column}, geometry~(B), wall $+$ gap $+$ ring: the same
donor and inner wall, followed by a compact inner disk (purple dotted) that truncates at
$0.08$~AU, with the region beyond it evacuated and the outer dust forming
a cool, optically-thin crystalline ring (teal dotted), which supplies the $15$--$21~\mu$m
photometric excess. In both panels the total model (thick blue solid) is the sum of the donor
photosphere (orange dashed) and the irradiated inner wall of large grains (red dot-dashed), which
reproduces the $5$--$15~\mu$m MIRI/LRS continuum, plus the outer component. Filled black circles
are the observed MIRI photometry (not donor-subtracted) with variability-inflated errors; the
open downward triangle is the F2550W $3\sigma$ upper limit, which both models respect.}
\label{fig:diad}
\end{figure*}

%***********
\subsection{Irradiated circumbinary-disk model}
\label{sec:diad}

Taking the thermal interpretation, the blackbody fit of \S\ref{sec:continuum} 
places the emitting material outside the binary. Equating the stellar-subtracted continuum to an
optically-thick blackbody at $T=764$~K and $D = 1.72$~kpc gives a luminosity
$L \simeq 8\times10^{30}~{\rm erg\,s^{-1}}$ and an emitting-area radius
$R_{\rm BB}\equiv[L/(4\pi\sigma T^4)]^{1/2}\simeq2.6\,R_\odot$, or $\simeq1.05\,a$ in units of the
binary separation $a = 2.5\,R_\odot$. $R_{\rm BB}$ is an emitting-area radius and hence a
lower limit on the physical radius: a disk subtending solid angle $\Omega$ lies at
$r = R_{\rm BB}(4\pi/\Omega)^{1/2}$ \citep{GandhiGX339}, so a covering fraction of
$\lesssim38\%$ places any dust beyond the tidal-truncation radius $\simeq1.7\,a$
\citep{ArtymowiczLubow1994}.

The same temperature also fixes the distance of the dust from its heat source. For an illuminating luminosity $L_\star + L_{\rm acc} \simeq 0.06\,L_\odot$ (see below), a directly irradiated surface of large grains reaches $764$~K at $r = [L/(4\pi\sigma T^4)]^{1/2} \simeq 5.6\,a$, and small grains radiating isotropically at $\simeq2.8\,a$. 
Since $r \propto L^{1/2}T^{-2}$, a $30\%$ error in $L$ moves these by only $14\%$. The two arguments are mutually consistent: at $2.8$--$5.6\,a$ the emitting area $R_{\rm BB}$ implies a covering fraction of $4$--$14\%$, readily supplied by a tall inner rim. Both place the dust well outside the tidal-truncation radius, with no assumption about its geometry.

\begin{deluxetable*}{lll}
\tablecolumns{3}% sets \pt@ncol so \cutinhead spans (and centres across) all three columns
\tablecaption{Representative DIAD model parameters for XTE~J1118+480: the wall$+$disk geometry~(A)
and the wall$+$gap$+$ring geometry~(B)\label{tab:diad}}
\tablehead{\colhead{Parameter} & \colhead{(A) Wall $+$ disk} & \colhead{(B) Wall $+$ gap $+$ ring}}
\startdata
\cutinhead{Donor star (common to both)}
$T_{\rm eff}$ & \multicolumn{2}{c}{$4100$~K} \\
$R_S$\tablenotemark{a} & \multicolumn{2}{c}{$0.46~R_\odot$} \\
Distance & \multicolumn{2}{c}{$1720$~pc} \\
\cutinhead{Inner wall / circumbinary rim (common to both)}
$T_{\rm wall}$ & \multicolumn{2}{c}{$800$~K} \\
$R_{\rm in}$ & \multicolumn{2}{c}{$0.065$~AU $\;(5.6\,a)$ } \\
Height & \multicolumn{2}{c}{$4.2\,H_{\rm p}$} \\
$a_{\rm min}$--$a_{\rm max}$ & \multicolumn{2}{c}{$1$--$100~\mu$m} \\
Composition & \multicolumn{2}{c}{amorphous pyroxene (Mg50)} \\
\cutinhead{Outer component (differs between geometries)}
Structure & compact optically-thick disk & compact inner disk $+$ gap $+$ optically-thin ring \\
Gap & \nodata & $0.08$--$0.30$~AU \\
Outer dust & disk, $0.065$--$0.10$~AU & disk, $0.065$--$0.08$~AU; ring, $0.30$--$0.45$~AU $\,(26$--$38\,a)$ \\
Optical depth & $\gg1$ (thick) & $\gg1$ (inner disk); $\tau_{10}\approx0.0046$ (ring) \\
$a_{\rm max}$ (outer) & $100~\mu$m & $0.25~\mu$m \\
Composition (outer) & amorphous silicate & crystalline (forsterite, enstatite) $+$ amorphous \\
Viscous $\dot{M}$ & $3\times10^{-11}~M_\odot~{\rm yr^{-1}}$ & $3\times10^{-11}~M_\odot~{\rm yr^{-1}}$ (inner disk) \\
\cutinhead{Geometry (common to both)}
$\cos i$ & \multicolumn{2}{c}{$0.28\;(i\simeq74^\circ)$} \\
\enddata
\tablecomments{These are representative parameters found by exploring the model space by hand, not the
result of a formal fit (\S\ref{sec:diad}); the two geometries share the same inner wall and differ only
in the outer component. Binary separation $a = 2.5~R_\odot$. The wall height is given in units of the gas pressure scale height at the wall, $H_{\rm p}=c_{\rm s}/\Omega$, where $c_{\rm s}=(kT/\mu m_{\rm H})^{1/2}$ is the isothermal sound speed for a mean molecular weight $\mu=2.34$ and $\Omega$ is the Keplerian angular frequency about the total binary mass. The donor photosphere is computed within DIAD
from the adopted stellar parameters \citep{Cherepashchuk2019}. In geometry~(A) the disk radius and viscous
accretion rate are weakly constrained and degenerate, and are not treated as measured quantities.}
\tablenotetext{a}{$R_S$ is the DIAD input describing the central illuminator, and is the
luminosity-equivalent sphere radius $\sqrt{L_\star/4\pi\sigma T_{\rm eff}^4}$ derived from the donor
template pinned to the contemporaneous optical monitoring (\S\ref{sec:stellar_sub});
$L_\star\simeq0.054~L_\odot$. }
%It is not the donor's geometric radius, which is set by its Roche lobe: the donor fills its lobe, is tidally distorted, and therefore has no single geometric radius. Likewise the DIAD central mass is the total binary mass.}
\end{deluxetable*}

A single-temperature fit does not establish whether a disk
in radiative and hydrostatic equilibrium, with realistic grain opacities, admits such a
configuration. To test that stronger claim we adopt the
D'Alessio Irradiated Accretion Disk (DIAD) code \citep{Dalessio1998, Dalessio2006}, which
solves self-consistently for the hydrostatic vertical structure of a viscously heated, externally
irradiated disk containing two grain populations, small ISM-like grains in the illuminated upper
layers and larger grains settled toward the mid-plane, each with a size distribution
$\propto a^{q}$ between $a_{\rm min}$ and $a_{\rm max}$. Developed for protoplanetary disks, this model
truncates the disk at the dust-destruction ``wall'' \citep{Natta2001, Dullemond2001} and is
routinely applied to disks with cleared inner regions
\citep[pre-transitional disks;][]{espaillat2007}.

The central illuminator is the inner binary system, which is treated as a single source: to the donor
photosphere ($L_\star\simeq0.054\,L_\odot$) we add the luminosity of the quiescent inner
accretion flow, estimated from its ultraviolet and X-ray (e.g. \citealt{McClintock2003}) emission as
$L_{\rm acc}\simeq2\times10^{31}~{\rm erg\,s^{-1}}\simeq0.005\,L_\odot$, supplied through DIAD's
accretion-rate parameter. The donor therefore dominates the irradiation by roughly an order of
magnitude. Because the LRS spectrum and photometry are
non-simultaneous, we do not attempt a rigorous joint fit; we instead vary the model parameters by
hand to identify physically plausible configurations that reproduce the main features of the SED.
%The models presented here are representative members of a family of acceptable solutions, not unique best fits. 
Representative parameters are listed in Table~\ref{tab:diad}.
The two geometries are sketched in the upper panels of Figure~\ref{fig:diad}.

The inner wall consists of warm dust dominated by
large amorphous-pyroxene grains. It reproduces the $5$--$12~\mu$m LRS continuum and the
$15~\mu$m photometric point. The predominance of large grains renders the wall emission nearly featureless, explaining why the LRS lacks a $9.7~\mu$m amorphous-silicate band (which would otherwise be produced by sub-micron grains; \citealt{Dalessio2006}).  Rather than a filled disk, a tall, directly-illuminated wall emits from a narrow bright rim. This geometry reproduces the observed flux density at $R_{\rm in} \simeq 0.065$~AU $\simeq 5.6\,a$, as the energy-balance estimate above requires, which locates the dust well outside the tidal-truncation radius. The dust-destruction wall therefore lies within the range of radii where a circumbinary disk
is dynamically stable.

Geometry~(A) serves as our baseline and represents the standard DIAD configuration: a continuous, irradiated wall plus a disk (Figure~\ref{fig:diad}, left). An optically thick viscous disk that fills the region immediately outside the wall successfully reproduces the LRS continuum and respects the $25.5~\mu$m upper limit; however, it fails to produce the peaked $15$--$21~\mu$m excess. Because a viscous disk is hottest at its inner edge, any disk bright or radially extended enough to produce that excess would overproduce long-wavelength emission and violate the $25.5~\mu$m limit. Conversely, a disk compact enough to respect the upper limit under-estimates the measured $18~\mu$m flux density.

Geometry~(B) is specifically motivated by the $18~\mu$m photometric point (Figure~\ref{fig:diad}, right). Producing a peaked excess at this wavelength without introducing an 
accompanying $9.7~\mu$m silicate band requires dust that is both cooler and optically thin. This condition can be achieved by truncating the disk at $0.08$~AU and evacuating the region beyond it, forcing the outer dust into a thin ring. A mixture of crystalline forsterite, enstatite, and amorphous silicates then generates a broad $18$--$24~\mu$m complex that reproduces the observed excess while remaining safely below the F2550W limit.

Both geometries successfully reproduce the LRS continuum and match the three photometric points within their variability-inflated uncertainties, all while respecting the $25.5~\mu$m upper limit. Given the non-simultaneous nature of the current data, neither model is statistically preferred; only a bright, radially extended disk can be firmly ruled out. The extra complexity introduced in Geometry~(B) is motivated entirely by the $18~\mu$m point, which exceeds its neighbors by less than
$1\sigma$ and may reflect variability between the non-simultaneous filters. A single-epoch MIRI/MRS spectrum would test whether this feature is real, and would also
resolve the marginal feature at $7.47~\mu$m; a confirmed line detection would require a thermal
component, since synchrotron produces no lines.

Following \citet{Muno2006}, the inferred optically thick dust column implies a dust mass of approximately $10^{22}$~g and a total circumbinary mass of order $10^{24}$~g assuming an interstellar gas-to-dust ratio. Such a circumbinary mass reservoir could leave a dynamical signature. XTE~J1118+480 exhibits an orbital-period decay of $\dot{P} = -1.90\pm0.57~{\rm ms\,yr^{-1}}$ \citep{GonzalezHernandez2014}, which is far faster than predicted by gravitational radiation and not easily explained by magnetic braking. Instead, a resonant tidal torque from circumbinary material could efficiently extract angular momentum \citep{Taam2001, ChenLi2015}. However, as already argued by \cite{Muno2006}, this inferred mass is insufficient to alter the system's orbital angular momentum evolution unless it has been underestimated by 4 to 5 orders of magnitude.

%%%%%%%%%%%%%%%%%%%%%
\section{Summary and Conclusions}
\label{sec:conclusions}

We have presented JWST/MIRI imaging (F1500W, F1800W, F2100W, and F2550W) and low-resolution
spectroscopy of the quiescent black hole X-ray binary XTE~J1118+480, obtained under program
GO~3832 as the companion to the A0620$-$00 study of \citet{Zuo2025}, together with
contemporaneous optical monitoring and a radio constraint. Our main results are as follows.

\begin{enumerate}

\item The stellar-subtracted $5$--$12~\mu$m LRS continuum is curved: a single power law is rejected relative to a single-temperature blackbody
($\Delta\chi^{2}=71$ at equal degrees of freedom; $T = 764\pm9$~K). This distinguishes
XTE~J1118+480 from A0620$-$00, whose stellar-subtracted spectrum is a single power law over the
same range.
 
\item Blackbody, broken-power-law, and single self-absorbed-synchrotron forms reproduce the
LRS curvature comparably well, so the LRS shape alone does not isolate the emission mechanism.
None of them, extrapolated, reproduces the MIRI photometry at $15$, $18$, and $21~\mu$m, which
lies $30$--$90\%$ above every extrapolation and 
persists at $\gtrsim 2.5\sigma$ after
a conservative allowance for the imaging scatter. An additional component is therefore suggested in every
interpretation. A single partially self-absorbed spectrum breaking near $6~\mu$m can reach the
photometry, but only at the expense of the LRS curvature and the $25.5~\mu$m limit. Its shape, however, is not well established, as the photometry was obtained
$2.6$~hr before the LRS. 
%The second synchrotron zone and the outer dust component of the disk models both rest on these points.
 
\item Modulo the lack of strict simultaneity, two self-absorbed synchrotron zones reproduce the full SED, with turnovers at $6.2$ and $17.5~\mu$m and, under equipartition, $B\simeq3.5\times10^{4}$ and $1.1$--$1.3\times10^{4}$~G on scales of $14$ and $29$--$37\,r_{\rm g}$. This scenario, if correct, does not resemble a steady, partially self-absorbed compact jet; a radio-quiet flickering jet, launching discrete ejections intermittently \citep{Gallo2019, dePolo2022}, remains possible.
 
\item Interpreted instead as thermal emission from warm dust irradiated by the binary, the
$764$~K continuum constrains the location of the dust: energy balance places it at $\simeq3$--$6$
binary separations, well outside the tidal-truncation radius of $1.7\,a$, independent of its
geometry.
Representative self-consistent irradiated-disk (DIAD) models reproduce the principal
features of the $5$--$25~\mu$m SED with a directly-illuminated dust wall at that distance, accounting for
the LRS continuum, and cooler dust farther out accounting for the $15$--$21~\mu$m excess. We
therefore favor thermal emission from a circumbinary dust disk, while noting that a
sub-dominant synchrotron contribution is neither required nor excluded. 
 
\end{enumerate}

Among the quiescent black hole X-ray binaries observed with JWST to date at $L_{\rm X}\lesssim10^{-8}\,L_{\rm Edd}$, XTE~J1118+480 is the first
whose mid-infrared excess favors a circumbinary dust disk, in contrast to the outflowing wind
inferred for A0620$-$00 \citep{Zuo2025}. The two systems are among the few BH-LMXBs faint enough for such a structure to be
detectable above the accretion flow, and they differ: the spectrum of
A0620$-$00 is a power law and varies on minute time-scales, whereas that of XTE~J1118+480 is
curved and steady. Whether the difference is intrinsic, or reflects the state of each system at
the time of observation, is not well established by these data; we note that the same disk modeling applied
to the A0620$-$00 spectrum is warranted and defer it to future work. If confirmed, such a disk
would supply the reservoir of cool circumbinary material long invoked in the evolution of compact
binaries. A definitive test will require a single-epoch MIRI/MRS spectrum, sensitive to the gas
emission lines that should accompany the dust.

\begin{acknowledgments}
This work is based on observations made with the NASA/ESA/CSA James Webb Space Telescope.
The data were obtained from the Mikulski Archive for Space Telescopes at the Space Telescope
Science Institute, which is operated by the Association of Universities for Research in
Astronomy, Inc., under NASA contract NAS 5-03127 for JWST. These observations are associated with program \#3832. Support for program \#3832 was provided by NASA through a grant from the Space Telescope Science Institute, which is operated by the Association of Universities for Research in Astronomy, Inc., under NASA contract NAS 5-03127. The JWST data presented in this article can be
accessed via this link: \url{https://doi.org/10.17909/v5w6-0h77}.
The National Radio Astronomy Observatory is a facility of the National Science Foundation
operated under cooperative agreement by Associated Universities, Inc. This work makes use of
observations from the Las Cumbres Observatory global telescope network.
R.M.P.\ acknowledges support from the National Science Foundation under grant No.\ 2511711.
D.M.R.\ is supported by Tamkeen under the NYU Abu Dhabi Research Institute grant CASS.
Anthropic's Claude (version Opus 5) was utilized during the 
preparation of this manuscript. Specifically, the tool was used to improve 
the clarity and grammar of the text. 
The authors reviewed, verified, and accept full responsibility for all 
final text.
\end{acknowledgments}

\facilities{JWST (MIRI), VLA, LCO}
 
\software{jwst \citep{Bushouse2024}, photutils \citep{Bradley2024}, astropy
\citep{astropy2013, astropy2018, astropy2022}, CASA \citep{casateam}, dynesty
\citep{Speagle2020}, NumPy \citep{Harris2020}, SciPy \citep{Virtanen2020}, Matplotlib
\citep{Hunter2007}, DIAD \citep{Dalessio1998, Dalessio2006}}

\clearpage
\appendix

\section{Las Cumbres Observatory monitoring}
\label{app:lco}
\begin{figure}
\begin{center}
\includegraphics[width=0.65\linewidth]{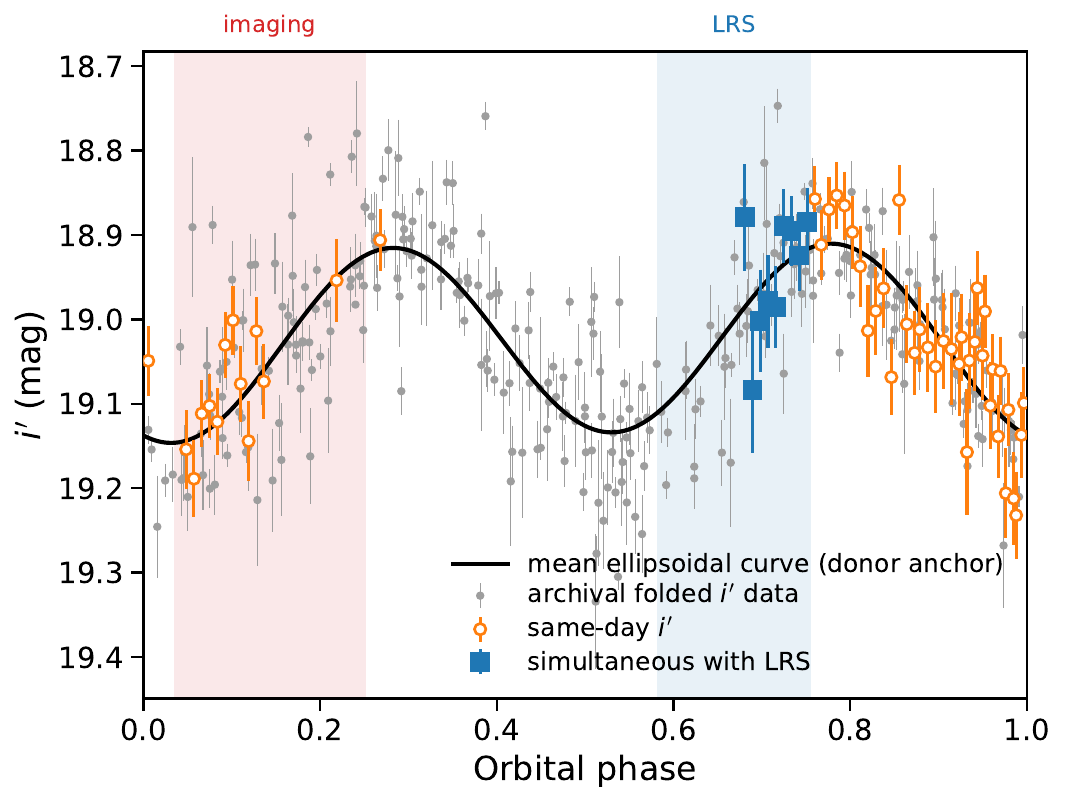}
\end{center}
\caption{Joint Fourier (ellipsoidal-modulation) fit to the phase-folded ground-based SDSS $i'$
monitoring of XTE~J1118+480. Gray circles are archival photometry; open orange circles are the
same-day-with-JWST monitoring used to anchor the donor template, and filled blue squares are the
nine points obtained during the LRS exposure itself. The orbital phase is computed from the \citet{Cherepashchuk2019} ephemeris.
Shaded bands indicate the phase coverage of the JWST/MIRI imaging (orange) and LRS (blue)
observation windows.}
\label{fig:donor_lc}
\end{figure}

XTE J1118+480 has been observed roughly weekly with the 2-m and 1-m robotic telescopes of the Las Cumbres Observatory (LCO), as part of an optical monitoring campaign of $\sim 50$ LMXBs \citep[][]{Lewis2008}. For the purposes of this work, we analyze archival data in SDSS $i^{\prime}$-band from 2011 May to 2026 January (a total of 379 detections with errors $< 0.1$ mag), and long-term data in Bessell $B$, $V$ and $R$ (from 2011) and SDSS $g^{\prime}$, $r^{\prime}$ and $z_{\rm s}$ (from 2020), each including epochs within $\pm$ one day of the JWST observation, which we use to check the donor normalization (\S\ref{sec:stellar_sub}). The images and calibration files, downloaded from the LCO archive \footnote{\url{https://archive.lco.global}}, were processed and analyzed using the X-ray Binary New Early Warning System (\textsc{XB-NEWS}), a real-time data analysis pipeline \citep[e.g.][]{Russell2019, Goodwin2020, Saikia2026}. The pipeline performed standard data reduction using LCO calibration files, several quality control steps, computed an astrometric solution for each image using positions from the Gaia DR2 catalogue\footnote{\url{https://www.cosmos.esa.int/web/gaia/dr2}}, and achieved flux calibration using aperture photometry of all the stars in each image, solving for zero-point calibrations between epochs \citep{Bramich2012}, adopting an extended version of the ATLAS All-sky Stellar Reference Catalogue \citep[ATLAS-REFCAT2;][]{Tonry2018}, and using multi-aperture photometry \citep[][]{Stetson1990} on the target in each reduced image. The orbital phase-resolved $i^{\prime}$-band light curve is shown in Figure~\ref{fig:donor_lc}.
Rapid flares lasting a few minutes are superimposed on the ellipsoidal modulation of
XTE~J1118+480 in quiescence \citep{Shahbaz2005}, but over 2011--2026 the $i'$ photometry scatters
about the ellipsoidal curve by only $0.06$~mag (robust rms), against a median photometric error of
$0.03$~mag, and only $1\%$ of points lie more than $3\sigma$ above it, by at most $0.25$~mag.
There is thus no sustained brightening of the kind that defines the active states of A0620$-$00
\citep{Cantrell2008}, which justifies anchoring the donor to the mean ellipsoidal curve
(\S\ref{sec:stellar_sub}).

\section{MIRI data reduction and donor subtraction}
\label{app:reduction}

\emph{Imaging.} We begin from the Stage~3 \texttt{i2d} products for each filter. The
background is estimated with the \texttt{MMMBackground} method in \texttt{photutils}, using a
mode estimator with a $3\sigma$ clipping threshold, and source detection is performed with
\texttt{DAOStarFinder}. The source aperture is a circular region of radius $r$ corresponding to
the 80\% encircled energy fraction (EEF), with the background estimated in a concentric annulus
of inner radius $2r$ and outer radius $3r$. Photometric uncertainties are the standard deviation
of the flux densities measured from 100 randomly placed apertures of radius $r$ within a source-free
region spanning $3r$ to $10r$. For the F2550W upper limit the noise is computed with apertures
of radius corresponding to the 50\% EEF, since the source is undetected at 80\% EEF; the
limit is corrected for the flux density outside that aperture.

\emph{Photometric variability.} F1500W and F1800W each comprise three dithers; F2100W
comprises twelve, which we combine into three groups of four because the source is marginally
detected ($\mathrm{S/N}\lesssim2$) in individual F2100W dithers. Starting from the Level~2
\texttt{cal} products, each dither or group is processed with \texttt{SkyMatchStep}
(\texttt{skymethod=global+match}), which estimates a common baseline sky across all images and
then applies additive corrections minimising sky-level mismatches between overlapping pairs, and
subsequently with \texttt{Image3Pipeline} with the \texttt{skymatch} step disabled. Aperture
photometry follows the prescription above.

\emph{LRS.} The observation consists of two nod positions with 20 integrations each.
Starting from the Level~1 \texttt{rateints} files, each nod is processed through
\texttt{Spec2Pipeline} using the opposite nod as background, producing a \texttt{calints} cube of
20 per-integration images; \texttt{Extract1dStep} is applied to each, yielding one spectrum per
integration. For the variability search we integrate each spectrum within three wavelength bands
($5$--$7$, $7$--$9$, $9$--$11~\mu$m), propagating the per-pixel \texttt{FLUX\_ERROR} in
quadrature, and normalise each nod by its own mean so that the nod-to-nod offset does not enter.

\emph{The science spectrum and its uncertainties.} The spectrum fitted in
\S\ref{sec:continuum} is the mean of the two nods, restricted to $5.06$--$11.99~\mu$m, where
both extractions agree; below $5~\mu$m the LRS throughput falls steeply and the two nods diverge
by tens of per cent, so those data are discarded. Because the nods are independent measurements
of the same spectrum, their disagreement calibrates the uncertainties without reference to the
pipeline error model. Over $4.8$--$12.2~\mu$m the difference between them gives
$\chi^{2} = 5661$ for $255$ points, where the quoted errors predict $\approx255$; the same
excess appears in the point-to-point scatter within each nod. We therefore rescale the pipeline
uncertainty at each wavelength by the ratio of the observed nod-to-nod scatter to that expected
from the quoted errors, computed in a running window of $41$ samples and floored at unity. The
resulting factor is $\approx4.4$ at $5.5~\mu$m, falling to $\approx1.6$ at $11~\mu$m, with a
median of $2.1$: the pipeline uncertainties are least reliable at the blue end, where the source
is brightest and the photon noise smallest, so that calibration systematics dominate the error
budget. The uncertainty on the donor normalisation (below) is added in quadrature to the
rescaled value at every wavelength.

\begin{figure}
\begin{center}
\includegraphics[width=0.65\linewidth]{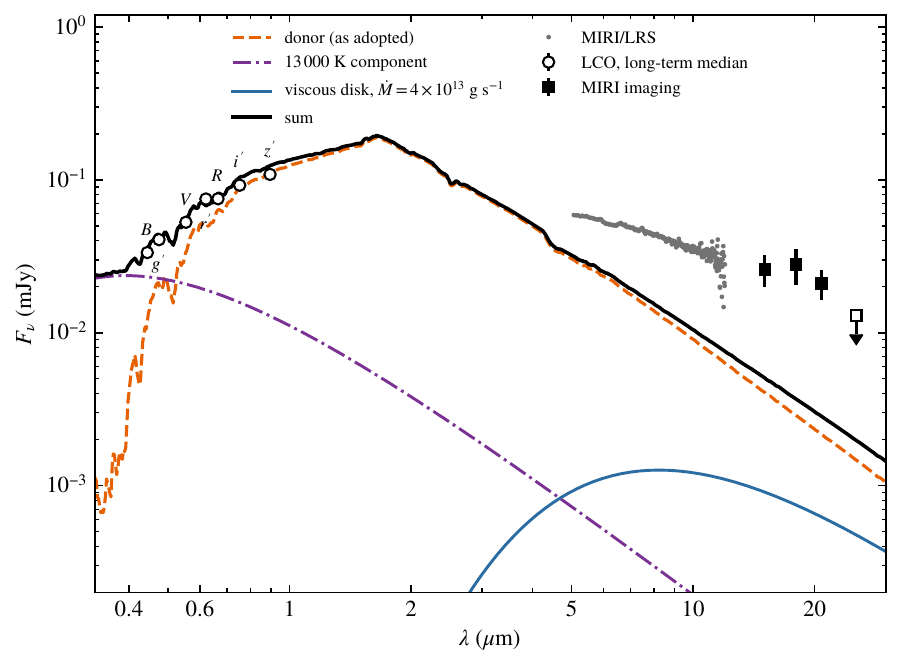}
\end{center}
\caption{Optical to mid-infrared SED of XTE~J1118+480. Open circles: long-term median LCO
photometry, de-reddened for $A_V=0.05$, with a $5\%$ error floor. Orange dashed: the BT-Settl
donor template at the normalization adopted in the analysis ($91\%$ of the total $i'$ flux density).
Dash-dotted: a $13\,000$~K blackbody representing the optical/UV component of
\citet{McClintock2003}, fitted to the optical residuals ($R\simeq2.3\times10^{9}$~cm). Blue: a viscously heated disk with $\dot M=4\times10^{13}$~g~s$^{-1}$ \citep{Muno2006},
extending from $\sim2\times10^{4}\,r_{\rm g}$ to $0.9\,R_{L1}$. Black: the sum. Gray points and black squares: total MIRI/LRS and imaging flux densities.
The two non-stellar components together contribute $\lesssim6\%$ of the mid-infrared excess.}
\label{fig:optical_sed}
\end{figure}

\emph{Donor subtraction.} The ellipsoidal modulation in each band is represented by
\begin{equation}
m_b(\varphi) = C_b + a_1\cos(2\pi\varphi) + b_1\sin(2\pi\varphi) + a_2\cos(4\pi\varphi) + b_2\sin(4\pi\varphi),
\end{equation}
with $\varphi$ the orbital phase from the \citet{Cherepashchuk2019} ephemeris and the
coefficients determined by weighted least-squares to the same-day $i'$ monitoring, which spans
two days and twelve orbital cycles and therefore determines the curve over the full phase range.
We use $i'$ alone for the normalization, as the only band with dense coverage contemporaneous
with the JWST visit; the long-term photometry in $B$, $V$, $R$, $g'$, $r'$ and $z'$ is used to
check it (\S\ref{sec:stellar_sub}; Figure~\ref{fig:optical_sed}). For
each JWST exposure window $[t_0,t_1]$ the time-averaged donor flux density is
\begin{equation}
\langle F_b\rangle = \frac{F_{0,\mathrm{AB}}}{\Delta t} \int_{t_0}^{t_1} 10^{-0.4\,m_b(\varphi(t))}\,dt,
\end{equation}
with $F_{0,\mathrm{AB}} = 3.631\times10^{6}~\mathrm{mJy}$, and the template is scaled by the
single window-dependent scalar
\begin{equation}
s_{\rm win} = f_\star\,\frac{\langle F_{i'}\rangle_{\rm obs}}{\langle F_{i'}\rangle_{\rm tmpl}},
\qquad
\langle F_\nu\rangle_{i'} = \frac{\int F_\nu(\lambda)\,T_{i'}(\lambda)\,d\lambda}{\int T_{i'}(\lambda)\,d\lambda},
\end{equation}
where $T_{i'}(\lambda)$ is the filter transmission and $f_\star=0.91$ is the fraction of the
$i'$ flux density attributed to the donor (\S\ref{sec:stellar_sub}). Because the imaging and the LRS fall on
opposite halves of the ellipsoidal cycle, $s_{\rm win}$ is evaluated separately for each
exposure rather than once for the visit. The scaled template is averaged over each filter
bandpass and over each LRS resolution element, rather than interpolated, and subtracted. The
uncertainty on $s_{\rm win}$ combines the formal covariance of the Fourier coefficients, the
dispersion of the monitoring data about the fitted curve within each window, and the $10\%$
systematic flux-density error associated with the K3V--M0V spectral-type range
\citep[cf.][]{Zuo2025}.

\section{VLA Data Reduction}
\label{app:vlareduction}
The VLA observed XTE~J1118+480 for 7.5 hours starting on 2024 April 12 00:30 UT in the VLA's C configuration (maximum baseline of 3.4 km), under project code 24A-274.  Observations were taken from 4--8 GHz (C-band) using the 3-bit {samplers}.  The primary flux calibrator  3C 286 was used to solve for antenna delays, to perform bandpass calibrations, and to set the flux density scale.  The secondary calibrator J1126+4516 was observed every nine minutes between science scans to solve for the time-dependent complex gain solutions.  We ultimately obtained 6h12m dwell time on XTE~J1118+480.

We downloaded calibrated data products from the VLA archive (using version 6.5.4.9 of the VLA calibration pipeline), and further data processing was performed using the Common Astronomy Software Applications package \citep[{\tt CASA;}][]{casateam} v6.5.4.  After visual inspection of the calibration solutions, and a small amount of additional flagging, we imaged our science field using the task {\tt tclean}.  During imaging, we used two Taylor terms {\tt nterms=2} to account for the wide fractional bandwidth, and we used Briggs weighting with {\tt robust=1.5} to balance sensitivity with minimzing sidelobes from nearby sources.  We measured the \textit{rms} noise level from a source-free region of sky near the known location of XTE~J1118+480, providing a thermal noise $\sigma_{\rm rms}=1.5$ $\mu$Jy bm$^{-1}$.  No statistically significant radio emission was detected from XTE~J1118+480, such that we place an upper limit at 3$\sigma_{\rm rms} < 4.5\, \mu$Jy.

\bibliographystyle{aasjournal}

\bibliography{bibliography_XRB.bib}

\end{document}